\documentclass[twocolumn]{aastex631}

\usepackage{CJKutf8} 
\usepackage{booktabs}

\newcommand\omc{$\omega$\,Cen}

\newcommand\colortwo{$\rm{C_{F275W,F336W,F435W}}$}

\newcommand\deltaone{$\rm{\Delta_{F336W,F814W}}$}
\newcommand\deltatwo{$\rm{\Delta_{C_{F275W,F336W,F435W}}}$}

\begin{document}
\begin{CJK*}{UTF8}{gbsn}
\title{oMEGACat. VIII. A Subpopulation Census of $\omega$ Centauri}

\correspondingauthor{Callie Clontz}
\email{clontz@mpia.de}

\author[0009-0005-8057-0031]{C. Clontz}
\affiliation{Department of Physics and Astronomy, University of Utah, Salt Lake City, UT 84112, USA}
\affiliation{Max-Planck-Institut für Astronomie, Königstuhl 17, D-69117 Heidelberg, Germany}

\author[0000-0003-0248-5470]{A. C. Seth}
\affiliation{Department of Physics and Astronomy, University of Utah, Salt Lake City, UT 84112, USA}

\author[0000-0003-2512-6892]{Z. Wang (王梓先)}
\affiliation{Department of Physics and Astronomy, University of Utah, Salt Lake City, UT 84112, USA}

\author[0000-0002-5844-4443]{M. H\"aberle}
\affiliation{European Southern Observatory, Karl-Schwarzschild-Strasse 2, 85748 Garching, Germany}
\affiliation{Max-Planck-Institut für Astronomie, Königstuhl 17, D-69117 Heidelberg, Germany}

\author[0000-0002-2941-4480]{M. S. Nitschai}
\affiliation{Max-Planck-Institut für Astronomie, Königstuhl 17, D-69117 Heidelberg, Germany}

\author[0000-0002-6922-2598]{N. Neumayer}
\affiliation{Max-Planck-Institut für Astronomie, Königstuhl 17, D-69117 Heidelberg, Germany}

\author[0000-0002-6922-2598]{P. J. Smith}
\affiliation{Max-Planck-Institut für Astronomie, Königstuhl 17, D-69117 Heidelberg, Germany}
\affiliation{Department of Physics and Astronomy, University of Heidelberg, Im Neuenheimer Feld 226, D-69120 Heidelberg, Germany}

\author[0000-0002-7547-6180]{M. Latour}
\affiliation{Institut für Astrophysik und Geophysik, Georg-August-Universität Göttingen, Friedrich-Hund-Platz 1, 37077 Göttingen, Germany}

\author[0000-0002-0160-7221]{A. Feldmeier-Krause}
\affiliation{Department of Astrophysics, University of Vienna, T\"urkenschanzstrasse 17, 1180 Wien, Austria}
\affiliation{Max-Planck-Institut für Astronomie, Königstuhl 17, D-69117 Heidelberg, Germany}

\author[0000-0001-9673-7397]{M. Libralato}
\affiliation{INAF, Osservatorio Astronomico di Padova, Vicolo dell’Osservatorio 5, Padova,I-35122, Italy}

\author[0000-0003-3858-637X]{A.\ Bellini}
\affiliation{Space Telescope Science Institute, 3700 San Martin drive, Baltimore, MD, 21218, USA}

\begin{abstract}
An understanding of the assembly history of the complex star cluster Omega Centauri has long been sought after, with many studies separating the stars on the color-magnitude diagram into multiple groupings across small magnitude ranges. Utilizing the oMEGACat combined astro-photometric and spectroscopic dataset we parse 14 subpopulations from the upper red-giant branch to below the main-sequence turnoff. We combine our results with previous works to estimate the age and age spread of each population. We find that the chemically enhanced (P2) populations are all $\sim$1 Gyr younger ($\sim11.6$ Gyr old) and have significantly higher intrinsic age spreads (0.6 Gyr) than the primordial (P1) populations ($\sim 12.6$ Gyr old, 0.3 Gyr spread), with the intermediate (Im) populations falling in between the two. Additionally, we connect for the first time the Chromosome Diagram to the two-stream age-metallicity relation, allowing us to link the P1 and P2 stars to the distinct star formation tracks, proposed to be in-situ and ex-situ contributions to the cluster's assembly. Our results are consistent with some suggested formation models and rule out others but no current model can explain all observed features of the subpopulations. 
\end{abstract}


\section{Introduction}
\label{sec:intro}

Omega Centauri (\omc) is well-known for its many unique features that distinguish it from typical globular clusters, including a large ($\sim$2 dex) spread in metallicity \citep{Johnson_2010, Johnson_2020, Nitschai_2023}, suggesting it is composed of multiple stellar populations. There exists a multitude of studies aiming to constrain the exact number of subpopulations and their unique characteristics, some focused on the red-giant branch (RGB) \citep{Marino_2012, Milone_2017a, Husser_2020}, some on the subgiant branch (SGB) \citep{Villanova_2007, Villanova_2014}, and others on the main-sequence (MS) \citep{Bellini_2017c, Latour_2021}. A few studies have focused on reproducing the full color-magnitude diagram (CMD) via fitting of isochrones and/or population synthesis models\citep{Joo_2013, Tailo_2016}. The different techniques for identifying subpopulations provide a varied number of groupings of stars depending on the features the targeted data highlight. While the results have a strong dependence on data type and quality, there are similarities among the findings.

In spectroscopic studies, typically three primary metallicity groups are identified: the Metal-Poor (MP), Metal-Intermediate (MI), and Metal-Rich (MR) \citep{Marino_2011, Tailo_2016, Bellini_2017c}. The MP population contains around half of the \omc\ stars, exhibits a oxygen-sodium anti-correlation, helium abundances of up to $Y$ = 0.40 \citep{Clontz_2025} and is lanthanum-poor, indicative of r-process nucleosynthesis.
The MI population contains 1/3 to 1/2 of the stars, displays a significant internal spread in elemental abundances, exhibits an extreme O-Na anti-correlation, and is lanthanum-rich, suggesting it has substantial s-process contributions to its enrichment. The MR population is a small anomalous group of stars, being significantly more iron rich than the rest of the groups with nearly all stars being Na-rich ($\sim 1$ dex) but having a large range of oxygen values (0 - 0.6 dex) \citep{Marino_2011}.

Photometry provides additional ways to parse subpopulations on the color-magnitude diagram (CMD) where varying the analyzed magnitude range (and thus stellar evolutionary phase) can highlight separate characteristic differences between the subpopulations. For example, subpopulations with different helium abundances separate out well on the MS while those with varying nitrogen abundance separate out well on the RGB.

The RGB is the best location on the CMD for parsing individual subpopulations as the intrinsic [Fe/H] variations manifest most strongly as color differences here. A common technique to distinguish photometric differences is to construct a pseudo-color-color diagram, referred to here as the ``chromosome diagram"(ChD). See \cite{Milone_2017a} and \cite{Nitschai_2023} as well as section \ref{sec:phase_II_initial_clustering} for details on its construction.
The ChD is information rich and has provided the basis for several studies of subpopulations \citep{Milone_2018, Milone_2020, Marino_2019}. Beyond individual subpopulations the ChD can also be used to determine light element abundances, such as helium, across metallicity \citep{Clontz_2025}. 

Additional probes into the formation channels of \omc\ and their respective subpopulations come from ages of individual stars or full populations which are typically constrained along the subgiant branch due to model dependencies being minimized in this region. \cite{Villanova_2007} measured relative ages for 80 SGB stars and distinguished 4 groups among them, while \cite{Joo_2013} used synthetic CMDs of 5 subgroups to model the populations, with age spreads up to 1.7 Gyr. A follow-up study by \cite{Villanova_2014} using twice as many stars as before was able to separate the SGB stars into 6 groupings based on age, finding an age spread of at least 2 Gyr. 

The age-metallicity relation (AMR) has been constrained by several studies \citep{Hilker_2004, Villanova_2007, Villanova_2014}, and was recently analyzed for a larger sample of stars ($\sim$7k) by \cite{Clontz_2024}. The increased resolution clearly showed the two-stream feature suggested in \cite{Villanova_2014}. These two discrete features in the AMR implies there are multiple formation pathways contributing to the formation of \omc. It is also worth noting the high likelihood that there are additional discrete sub-structures in the age-metallicity diagram which are not yet distinguishable due to the blurring effect of uncertainties on the ages and iron abundances. Additionally, the connection of these age-metallicity streams to the P1, Im, and P2 streams of the ChD is not straightforward, as the SGB ChD is less informative due to the subpopulation color range narrowing in this CMD region, especially in \colortwo. 

Earlier studies at focused on fainter magnitudes, unveiling a puzzling split main-sequence \citep{Anderson_2002, Norris_2004, Piotto_2005}. Metallicity measurements indicated the blue MS was more metal-rich than its red counterpart, which was later explained by an enhancement in helium in the bMS \citep{Dupree_2013, Karakas_2014, Milone_2020, Reddy_2020, Hema_2020}. Another study of the MS by \cite{Bellini_2017c} combined multiple photometric bands to identify distinct sequences. The five core groups they identified (the MSa, rMS, bMS, MSd, and MSe) were further parsed in a color-color space into 15 subpopulations. These groups were found to occupy distinct regions of the ChD of the main sequence. 

The presence of multiple stellar populations with a significant spread in ages and diverse chemical properties points to \omc\ being the remnant core of a disrupted dwarf galaxy \citep{Norris_1996,Bekki_2003,Hilker_2004,Johnson_2010,Villanova_2014}, especially since the successive starbursts with increasing enrichment observed are best facilitated by the larger host's ability to retain supernova ejecta. While many studies have linked \omc\ to the Gaia-Enceladus dwarf galaxy merger event \citep{Pfeffer_2021,Callingham_2022, Limberg_2022}, the debate remains open as to the cluster's previous host dwarf galaxy \citep{Majewski_2012, Myeong_2019,Forbes_2020}. 

In summary, while progress has been made in identifying the primary groups of \omc's subpopulations, the complex variations within these groups necessitate a combination of probes to allow a better-resolved picture of the cluster's evolutionary history. Subpopulation studies of \omc\ have been confined to small regions on the CMD where populations are separated into groups using a single probe (AMR, ChD). We present, in this work, subpopulations which are separated from the upper RGB, through the SGB, to below the main-sequence-turnoff (MSTO). This allows for the connecting of the ChD with the AMR and thus the connection of probes of different aspects of the cluster's assembly history. This work will also facilitate further spatial and kinematic studies and ultimately a clearer picture of the assembly of \omc. 

In Section \ref{sec:data} we briefly detail the dataset utilized in this study and in Section \ref{sec:methods} we outline our subpopulation parsing methodology in four phases: Sample Preparation, Initial Clustering, Iterative Propagation, and Informed Re-iteration and Final Results. In Section \ref{sec:results} we describe our results and cross-match them with the ages catalog of \cite{Clontz_2024} where after we highlight the connection of assembly probes. In Section \ref{sec:discussion} we discuss interpretations of our results and compare our results to the literature. Finally, in Section \ref{sec:conclusions} we summarize our conclusions and outline future works planned. 

\section{Data} 
\label{sec:data}

In this study, we utilize the oMEGACat - the combination of the The Hubble Space Telescope (HST) astro-photometric dataset \citep{Haeberle_2024} and the Multi-Unit Spectroscopic Explore (MUSE) spectroscopic dataset \citep{Nitschai_2023} for \omc.  The HST data were retrieved from the Mikulski Archive for Space Telescopes (MAST) at the Space Telescope Science Institute and are accessible via \dataset[doi: 10.17909/26qj-g090]{https://doi.org/10.17909/26qj-g090}. This catalog provides photometric data in six broadband HST filters, each with specific corrections applied for the differential reddening and instrumental effects. 
The MUSE data can be accessed via:\dataset[IOP Science]{https://content.cld.iop.org/journals/0004-637X/958/1/8/revision1/apjacf5dbt3_mrt.txt}. In the published MUSE catalog the metallicity is reported as [M/H], as it is measured via full-spectrum fitting. They also apply an atomic diffusion correction which ensures the reported metallicities reflect the birth abundance of each star \citep{Nitschai_2023}. More than 300,000 stars in the oMEGACat have measurements in all six photometric bands along with a reliable metallicity. We refer to the above two catalog papers for detailed data reduction procedures and the relevant data correction procedures.

To make direct comparisons with previous works more straightforward, we convert the $\rm{[M/H]}$ values to $\rm[Fe/H]$ using Equation 3 from \cite{Salaris_1993}:

\begin{equation}
    \rm{[Fe/H] = [M/H] - \log(0.638 \times 10^{[\alpha/Fe]} + 0.362)},
\end{equation}
assuming $\rm{[\alpha/Fe] = 0.3\,dex}$ \citep{Norris_1995, Johnson_2010}. The uncertainty in our [Fe/H] values thus depends on the uncertainty (and star-by-star variations) in the abundances of alpha elements, including oxygen. \cite{Marino_2012} showed oxygen can increase by $\sim 0.2$ dex from [Fe/H] = -1.9 to [Fe/H] = -1.05, which is a source of systematic uncertainty in our [Fe/H] estimates, although its contribution is similar to the median metallicity error (0.08 dex). Hereafter all mentions of the metallicity of a star or stellar population is a reference to the [Fe/H] value unless otherwise stated. 

\section{Subpopulation Parsing Methodology} 
\label{sec:methods}

\begin{figure*}[htb]
    \centering
    \includegraphics[width = \textwidth]{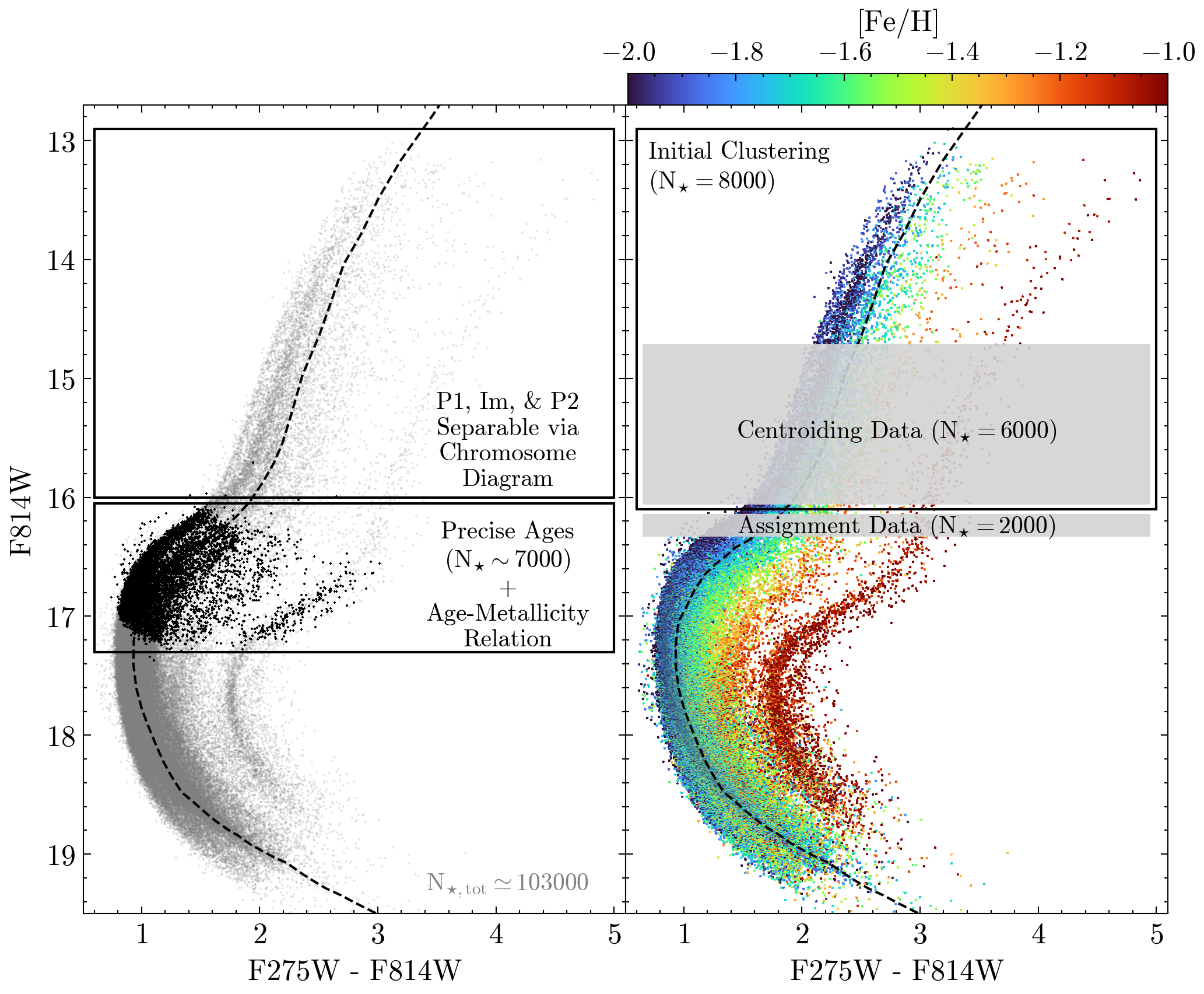}
    \caption{\textbf{Color-Magnitude Diagrams}:\textit{(both panels)} The black dotted line traces the overdense region of this plot when considering a subset of stars centered around the median metallicity ([Fe/H] $\sim$-1.7). \textit{(left panel)} The full subset of high-quality member stars is plotted with light grey markers. All SGB stars with age determinations are overplotted in black. The upper black box delineates the region where the RGB ChD is generated while the lower black box delineates where the age-metallicity relation is calculated. \textit{(right panel)} The full subset of high-quality member stars is plotted with individual markers colored by their [Fe/H] value, following the colorbar at the top of this panel. The black box delineates where the initial RGB clustering is performed. The upper grey shaded region shows the extent of the first propagation step centroiding sample while the lower grey shaded region shows the extent of the first propagation step assignment data. The scaling data is constituted by the combination of the centroiding and assignment data.}
    \label{fig:cmd_with_regions}
\end{figure*}

\begin{figure*}[htb]
    \centering
    \includegraphics[scale=.7]{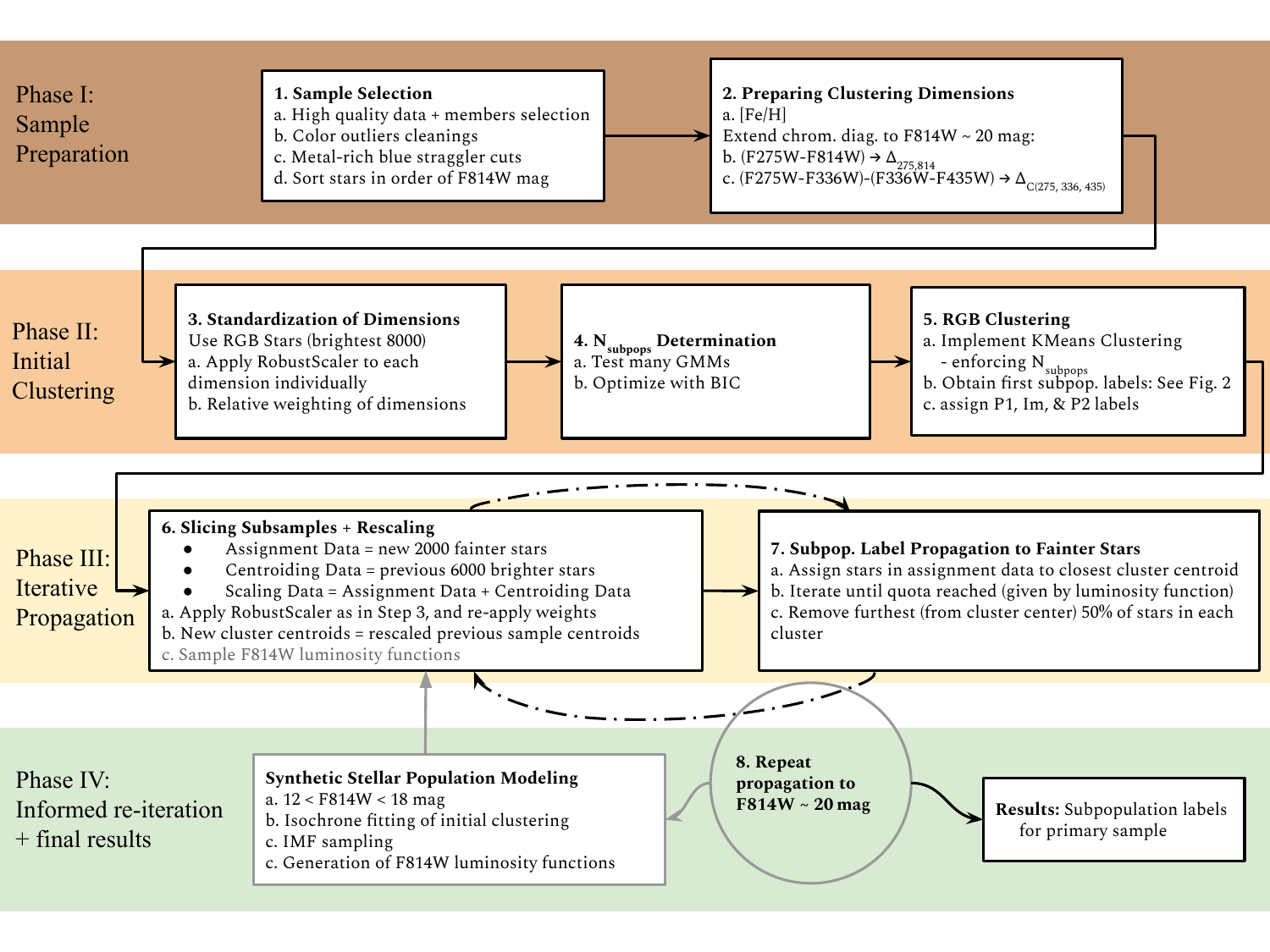}
    \caption{\textbf{Subpopulation Parsing Algorithm Flowchart}: The four phases of our subpopulation parsing algorithm are outlined vertically, top to bottom, with each phase being delineated by a horizontal colored band. Within each white box a bulleted list of grouped steps in the procedure is given. Black solid line arrows connect steps performed in order, and black dash-dot line arrows show steps performed iteratively. The grey lined boxes and arrows indicate steps which are only considered in our second-pass version of the subpopulation parsing algorithm.}
    \label{fig:flowchart}
\end{figure*}

A schematic of our clustering algorithm is shown in Figure \ref{fig:flowchart}. This flowchart outlines each of the steps taken to arrive at the final results; we discuss each of the phases of our procedure in the following subsections. In Appendix \ref{appendix:algorithm_variation_testing} we discuss the various methods which were tested to arrive at our final algorithm and how each choice was informed by the outcome of these tests.

\subsection{Phase I: Sample Preparation}
\label{sec:phaseI_sample_prep}

This study focuses on parsing individual continuous subpopulations from the upper RGB (F814W$\sim$13) to below the MSTO (F814W$\sim$20). Phase I: Sample Preparation, indicated by the brown band in Fig. \ref{fig:flowchart}, details the preparations taken to ensure clean, high-quality samples on which we perform our subpopulation analysis. 

\subsubsection{Sample Selection}
\label{sec:step_1_sample_selection}
This step includes all of the photometric and spectroscopic quality cuts applied to our data, as well as additional steps taken to remove outliers. 

To select only cluster members we exclude stars $> 3\sigma$ from the cluster velocity in both right-ascension and declination proper motion space (see \cite{Haeberle_2024} for details). Additionally, we apply radial velocity cuts using $RV_{probability}$ $>$0.9, given by the \cite{Nitschai_2023} catalog.

We apply additional criteria to keep only the most reliable data. These include a high-quality photometry flag as well as an upper cut on the $\chi^{2}$ weighted errors for each photometric filter we consider (both provided in the \cite{Haeberle_2024} catalog). Additionally, we select all stars in the range $-2.3<\rm{[Fe/H]}<-0.5$ and that have an [Fe/H] uncertainty of $<0.3$~dex. Lastly, we require the MUSE signal-to-noise ratio to be $>$ 10 and the Reliability parameter to be $>$ 0.95 (both provided in the \cite{Nitschai_2023} catalog). These combined cuts give 106,860 high-quality candidate stars located within the half-light radius of \omc. 

During the testing phase of our clustering methodology, we found that both blue and red color outliers in our color magnitude diagrams can affect the ability of the clustering algorithm to properly trace subpopulations to fainter magnitudes. This is due to these outliers having a notable effect on the calculation of the centroid of the subpopulation group, which causes the centroid to move, making it difficult to accurately propagate cluster labels to fainter magnitudes (details on this process are outlined in Section \ref{sec:step_7_propagation}). To correct for this, we performed a general cleaning of the $\rm{F275W-F814W}$ vs. F814W CMD by creating hand-drawn fiducial lines on both the blue and red extremes of the main locus of points, removing blue stragglers and binaries in the process. We also provide a conservative cleaning of the pseudo-color CMD ($\rm{C_{F275W,F336W,F435W}}$ vs. F814W). Together these cleanings remove a total of 363 stars. 

Through additional tests, we find there is a metal-rich blue straggler population, which also causes the centroids of our subpopulation centroids to migrate as we propagate to fainter magnitudes. These could not be removed via the previous red and blue fiducial line cleaning method given that they fall within the main locus of the more metal-poor MSTO stars. To isolate them, we separate the most metal-rich stars into three [Fe/H] bins and further remove the blue straggler populations from each using hand-drawn color-fiducial lines, similar to the process outlined above. This removes an additional 3,038 stars, leaving us with 103,459 well-measured stars for our initial sample. This sample is plotted in grey points in the left panel of Fig.\ref{fig:cmd_with_regions} and in the right panel each stars is colored its [Fe/H] value.

\subsubsection{Preparing Clustering Dimensions}
\label{sec:step_2_prep_clustering_dimensions}

This step includes the preparation of the dimensions in which the clustering and label propagation is performed. In addition to [Fe/H], we utilize the two dimensions of the ChD.

To track the populations across the full range of magnitude (12.9 $<$ F814W $<$ 19.7) we build upon methods used to construct the RGB ChD, outlined in several previous works \citep{Milone_2017a, Nitschai_2023, Clontz_2025}. These methods include verticalizing the bulk of the CMD sequences by using fiducial lines to create a pseudo-color-color (chromosome) diagram (\deltaone vs. \deltatwo). Here stellar populations with varying light element abundances separate well. Canonically, the x-axis is \deltaone,  formed from the verticalization of the F814W vs. F275W-F814W CMD, providing a long color-baseline and thus a proxy for metallicity and helium \citep{Clontz_2025}. The y-axis is typically \deltatwo, generated from the verticalization of the pseudo-color $\rm{C_{F275W,F336W,F435W}}$, which is the difference between two colors: (F275W-F336W)-(F336W-F435W). 

This ``magic trio" \citep{Milone_2012a, Piotto_2015} of filters highlights the separation of stars with varying abundances, due to the OH band falling in the F275W filter, the  NH band falling in the F336W filter and CN and CH falling in the F438W filter. In the RGB ChD of \omc\ (see Fig.\ref{fig:rgb_clustering}) there exists two distinct streams of stars extending from the top left to the bottom right, with a less populated middle stream falling in between the two. The lower stream is constituted of primordial (P1) populations which are helium and nitrogen-poor, while the upper stream contains chemically processed (P2) populations which are helium and nitrogen-enhanced. Those falling in the middle stream are considered Im populations and are thought to have intermediate abundances compared to P1 and P2. 

For this work, we need to generate the verticalized colors for the ChD axes from the upper RGB down past the MSTO (12.9 $<$ F814W $<$ 19.7). To do this we first create fiducial lines by isolating a sample of stars in a narrow metallicity range centered on the median ([Fe/H] $\sim$-1.7). We then hand-draw a line which traces the densest region of the metallicity sample along the CMD, then interpolate this line at each star's F814W magnitude. We calculate $\rm{\Delta_{F275W,F814W}}$ and $\rm{\Delta_{C_{F275W,F336W,F435W}}}$, by taking the difference between the interpolated fiducial line color and the color of each star. We then normalize the delta colors by the difference between the red and blue cleaning fiducial lines. These two delta colors, together with [Fe/H], form the 3-dimensional space in which we perform our subpopulation parsing analysis. We also note that because the SGB slope and MSTO magnitude are highly dependent on helium abundance and metallicity, extending the ChD prescription beyond the RGB in a way that is fair to all of the subpopulation CMD morphologies is quite tricky and results in an imperfect verticalization which we work to compensate for in future steps of our parsing algorithm.

\subsection{Phase II: Initial Clustering}
\label{sec:phase_II_initial_clustering}

Next we need to determine the number of subpopulations identifiable in our data and assign the first subpopulation labels to individual stars. To do so we turn to the RGB, where the clustered overdensities on the ChD, indicative of individual stellar populations, are known to separate best. Phase II, indicated by the orange band in Figure \ref{fig:flowchart}, details these steps. We start by selecting the brightest 8,000 stars (12.9 $<$ F814W $<$ 16.1), covering the main extent of the RGB, shown by the black box in the right panel of Fig.\ref{fig:cmd_with_regions}.

\subsubsection{Standardization of Dimensions}
\label{sec:step_3_standardization}

We then standardize our RGB sample in each of the three utilized dimensions (\deltaone, \deltatwo, and [Fe/H]). To do this, for each dimension we subtract the mean, which centers the data at zero. We then scale the $\rm{16^{th}}$ and $\rm{84^{th}}$ percentiles to unit variance. These steps allow us to equalize the weight each dimension is given and centers the data at the origin, which helps the clustering algorithms perform best. This method was found to be more reliable than other scaling methods due to its insensitivity to outliers. 

While scaling gives the same weight to each dimension, some data have more constraining power than others. This constraining power also changes with magnitude. Through empirical testing, we arrive at a set of weights that we find best help the clustering algorithm isolate overdensities and propagate membership labels in the combined ChD and [Fe/H] space. We up-weight the [Fe/H] dimension by scaling the values by 1.7. Similarly, we scale the \deltatwo values by 1.5. We find that up-weighting the [Fe/H] dimension by more than this causes poor performance on the main-sequence, but weighting it less than this causes poor clustering performance around the SGB and MSTO. Similarly, the \deltatwo axis becomes critical below the MSTO as the main diagnostic for subpopulation separation. While we did test changing the relative weights of each dimension as a function of magnitude, we found the best results when keeping the scalings consistent along the full CMD. 

\subsubsection{$N_{subpops}$ Determination}
\label{sec:step_4_n_subpops_determination}

In Step 4, we determine the number of distinguishable subpopulations in our data by passing the scaled and weighted data to a Gaussian Mixture Model \citep{scikit-learn}. We test the goodness of models with 1 to 25 clusters and for each we calculate the Bayesian Information Criterion value, which is minimized when the $\chi^2$ and number of model parameters are balanced. We find a minimum BIC at N$_{clusters}$ = 14. We chose this as our basis for our subpopulation sample but do not assert that this is the absolute number of subpopulations in \omc\ nor that each grouping is a pristine single stellar population. 

\subsubsection{RGB Clustering}
\label{sec:step_5_rgb_clustering}

\begin{figure*}[htb]
    \centering
    \includegraphics[width = \textwidth]{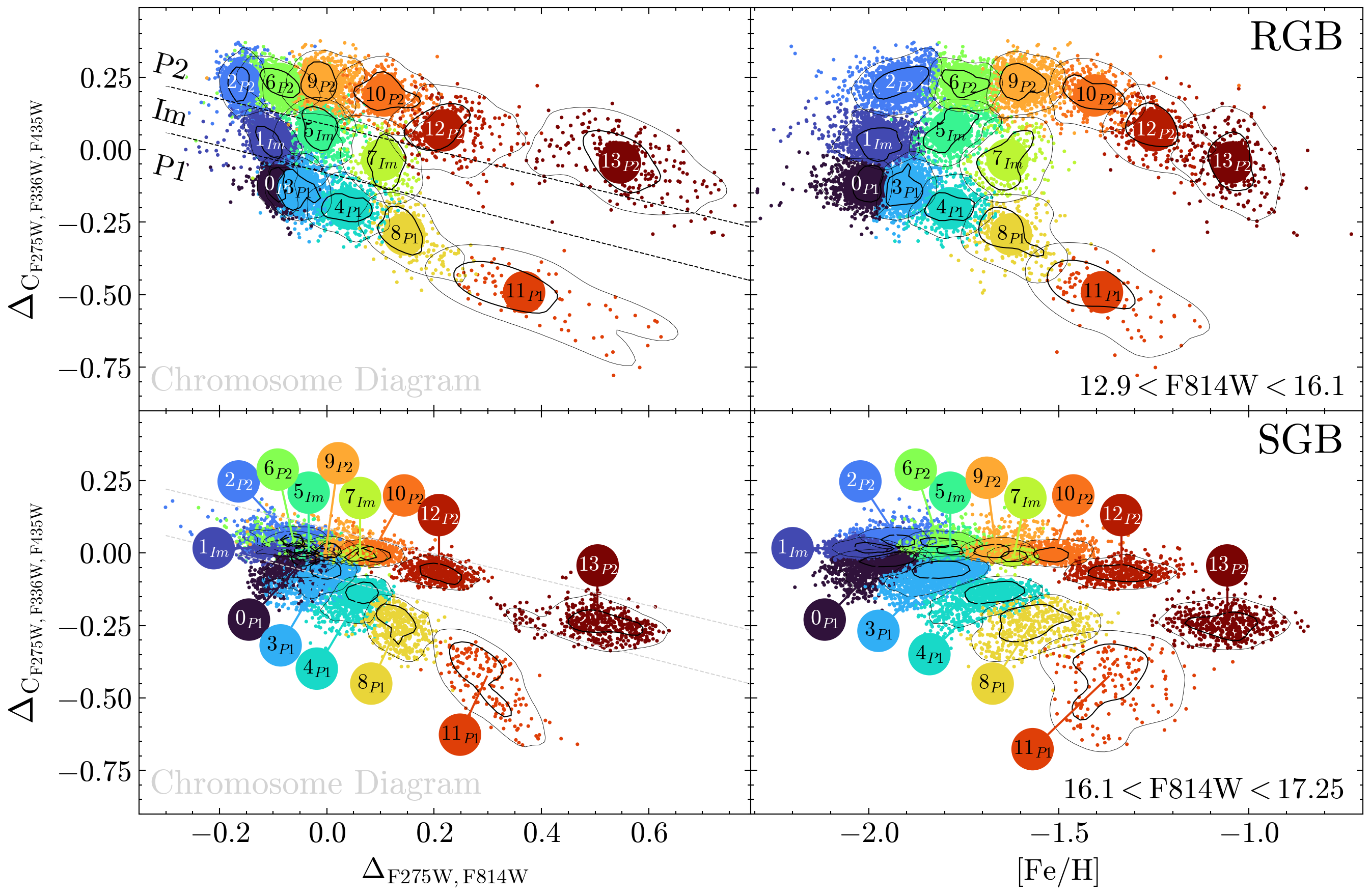}
    \caption{\textbf{Subpopulation Parsing:}\textit{(all panels)} Each small marker point is a single star, colored by its subpopulation label. The large colored circle marks the associated cluster and within each is the annotation of the cluster number, in order of metallicity, as well as the ChD stream label, as a subscript. The 1- and 2-$\sigma$ density contours are shown in thick and thin black lines, respectively. \textit{(upper left)} The RGB chromosome map is constructed from the two delta-colors. The centroid of each subpopulation is given by the large colored markers. \textit{(upper right)} The pseudo-color vs. [Fe/H] space shows the distinction of the subpopulations in this space. \textit{(lower panels)} Same as upper panels except now for the SGB. Here, the large colored markers are offset to aid visibility and are connected to the centroid of each cluster via a solid line with the same color.}
    \label{fig:rgb_clustering}
\end{figure*}

With the number of subpopulations determined, we next need to calculate the centroids of each cluster in the three-dimensional space and assign subpopulation labels to each RGB star. In Step 5, we implement a K-Means \citep{scikit-learn} clustering algorithm using the RGB sample (8000 brightest stars) from Section \ref{sec:step_3_standardization} and the 14 subpopulations (determined in Section \ref{sec:step_4_n_subpops_determination}) as a prior. This routine assigns a number from 0-13 (in no particular order) to each star, designating its subpopulation label. We take the median metallicity of the RGB stars of each population and re-sort the subpopulation groups by [Fe/H] and  reassign their labels in ascending order such that the most metal-poor population is Pop. 0 and the most metal-rich is Pop. 13. In the upper panels of Fig. \ref{fig:rgb_clustering} we show the initial clustering for the RGB where each star is colored by its sorted population label. The upper left panel is the classic ChD, illustrating the discreteness of the subpopulations as clustered points in this space. We also show here the dividing lines for the P1, Im, and P2 populations, which follow the underdense regions between the three primary streams. In the upper right panel, the populations are plotted against [Fe/H], where they are further distinguished. In both panels, we also overplot the 1-$\sigma$ and 2-$\sigma$ density contours in thick and thin black lines, given by the kernel density estimation of each cluster. 

To facilitate comparisons between populations that reside in each of these streams, we use the subpopulation centroids on the RGB ChD (upper left panel of Fig. \ref{fig:rgb_clustering}) to assign P1 labels to Pops. 0, 3, 4, 8, and 11, Im labels to Pops. 1, 5, and 7, and P2 labels to Pops. 2, 6, 9, 10, 12, and 13. It is important to clarify that our P1, Im, and P2 populations are analogous to "lower stream", "mid-stream", and "upper-stream" definitions in the literature \citep{Marino_2019, Clontz_2025, Dondoglio_2025}. Similarly, previous studies of the ChDs of various clusters, such as \cite{Milone_2017a}, denote the lower stream metal poor population (our Pop. 0) as 1G (first-generation) and the mid- and upper-stream metal-poor stars (our Pops. 1 and 2) as 2G (second-generation) stars. Here we do not adopt this nomenclature due to it implying a formation timeline. 

\subsection{Phase III: Iterative Propagation}
\label{sec:phaseIII_iterative_propagation}

In Phase III, highlighted by the yellow band in Fig.\ref{fig:flowchart}, we select and scale our next sub-samples of stars without subpopulation labels and propagate the RGB subpopulation labels to fainter magnitude stars using an iterative approach. This is repeated until we reach F814W $\sim$19.7 mag.

\subsubsection{Slicing sub-samples \& Rescaling}
\label{sec:step_6_preparing_to_propagate}

In Step 6, we create several sub-samples of our labeled data, which we describe in detail below, that are important for aligning the labeled data with the next set of stars to receive population labels. 

To begin, we create a sample of 8000 stars (referred to as the ``scaling sample"), which contains the faintest 6000 stars from the previous sample (referred to as the ``centroiding sample", indicated by the red shaded region in Fig.\ref{fig:cmd_with_regions}) along with the next 2000 fainter stars needing population labels assigned (referred to as the ``assignment sample", indicated by the blue shaded region in Fig.\ref{fig:cmd_with_regions}). This means that for the first step we take the 6000 faintest RGB stars as centroiding sample. The choice regarding the number of stars contained within the centroiding sample is made through iterative testing of the performance of our clustering algorithm with the main priority of ensuring good number statistics while at the same time maintaining enough flexibility for the algorithm.

Due to the metallicity dependence of the MSTO magnitude and the varying SGB slope due to helium abundance variations, the verticalization of the \deltaone and \deltatwo dimensions is not consistent across the full CMD. However, for propagation algorithm to track the same populations to fainter magnitudes it relies on stars that belong to the same population to live in the same region of the ChD. At fainter magnitudes these population even migrate across the three-dimensional clustering space, meaning we need to make sure the algorithm can similarly migrate the cluster centroid for this subpopulation to keep trace of its constituents. To aid in direct mapping, we use the scaling sample to re-normalize the extent of the centroiding and assignment samples. To aid in tracking migrations of clusters at fainter magnitudes we recalculate the new re-scaling at each progressive step in magnitude. To do this, we follow the same process described for the RGB initial clustering, meankng we remove the mean of each dimension, scale the 1-$\sigma$ range to unit variance, and apply relative weights to each dimension. 

From the scaling sample, we extract the centroiding data and use it to re-calculate the centroids of each subpopulation in the 3-dimensional space using the newly rescaled dimensions, ensuring the previous cluster centroids have been effectively transformed into the newly rescaled space. These are the cluster centroids we use in Step \ref{sec:step_7_propagation} to assign labels to the assignment data. 

\subsubsection{Subpopulation Label Propagation}
\label{sec:step_7_propagation}

In the first iteration of Step 7, we allocate cluster labels to the stars in the ``assignment" data by finding, for each star, the closest subpopulation cluster centroid from the centroiding data and assigning that tag. After this is complete for all assignment data stars, we move on to Step 8, where we shift our samples down by 2000 stars and follow Steps 6 and 7 again, and then again. We iterate down to F814W = 19.7 to arrive at our "first pass" version of our subpopulation tagging. We then take these preliminary clusters as a template for a model of the luminosity functions, a process which is outlined in Phase IV. 

\subsection{Phase IV: Informed Re-iteration + Final Results}
\label{sec:phaseIV_reiterate_and_final_results}

Before beginning the second and final pass of the subpopulation cluster label propagation algorithm, we model the luminosity functions of the first-pass subpopulations to inform our label propagation. This modeling helps us account for the increasingly fainter MSTOs of the more metal-rich and/or older populations, resulting in the relative number of stars contained within each subpopulation not being consistent across various magnitude ranges.

To begin, we take the first-pass subpopulations and select representative isochrones from a grid of models adapted from the Dartmouth Stellar Evolution Database \citep{Dotter_2008} (see also \cite{Clontz_2024} for more details). The isochrone grid contains models covering four parameter ranges: 8 $<$ Age $< 14$ Gyr (0.25 increments), -2.5 $<$ [Fe/H] $<$ -0.5 (0.01 increments), [$\alpha$/Fe] = 0.2, 0.3, and 0.4, and Y = 0.245 and Y = 0.40 and has a built in consideration of the C+N+O vs. [Fe/H] relation for \omc. 

We chose the isochrone model which has the median metallicity of its constituent RGB stars, the median age of its constituent SGB stars (discussed further in Section \ref{subsec:sgb_ages}), [$\alpha$/Fe] = 0.3 (as assumed in \cite{Clontz_2024}), and a helium mass fractions (Y = 0.25 and Y = 0.40) based on constraints from \cite{Clontz_2025} where P1 populations are all assumed to have Y = 0.25 and all P2 populations are helium enhanced, with Y = 0.40. The intermediate population is shown to have slight helium enhancements but the coarseness of our helium model grid means we chose to represent these populations with Y = 0.25 models. Then, using each subpopulation isochrone model, we finely sample a Kroupa IMF \citep{Kroupa_2001} within the mass range given by the isochrones and create an F814W luminosity function. Because our isochrone models are only calculated down to F814W $\sim$18, this is the faint-end extent of our luminosity function constraints. Beyond this magnitude, we use the fraction of RGB stars contained in each subpopulation as the constraint for the number of stars assigned to each subpopulation, which is a reasonable approximation on the main-sequence. One caveat is that due to our stringent data quality cuts, we cannot expect the luminosity functions to perfectly follow those modeled from synthetic populations. This is especially true if any of our quality selections preferentially affect stars of a given metallicity. 

We then begin a second pass of our clustering algorithm. We keep everything from Steps 1 through Step 6 the same. We start our second pass at Step 7, where we begin to propagate the subpopulation assignments to fainter stars. Now we can determine, from our modeled luminosity functions, how many stars in a given magnitude range should belong to each subpopulation ($\rm{N_{exp,i}}$, where i is the subpopulation index). Using this information, we take each subpopulation centroid and we find the $\rm{N_{exp,i}}$ closest stars to it and assign those stars the relevant subpopulation label. Before moving on to the next subpopulation cluster centroid we remove stars which have a distance from their cluster centroid greater than the $\rm{50^{th}}$ percentile distance of all stars in that cluster. This helps us prevent outliers from causing our populations to migrate away from their native over-densities and still preserves the relative fraction of subpopulation constituents dictated by the luminosity function samplings in the previous step. We also note that we do not apply this step to the RGB part of our subpopulations, meaning our subpopulation luminosity functions are not fully exemplary of the models on the RGB. After we do this step for one subpopulation, we move to the next subpopulation cluster centroid and repeat until we have assigned $\rm{N_{exp,i}/2}$ stars to all subpopulations in this magnitude range. 

\section{Subpopulation Results} 
\label{sec:results} 

\begin{figure*}[htb]
    \centering
    \includegraphics[width = \textwidth]{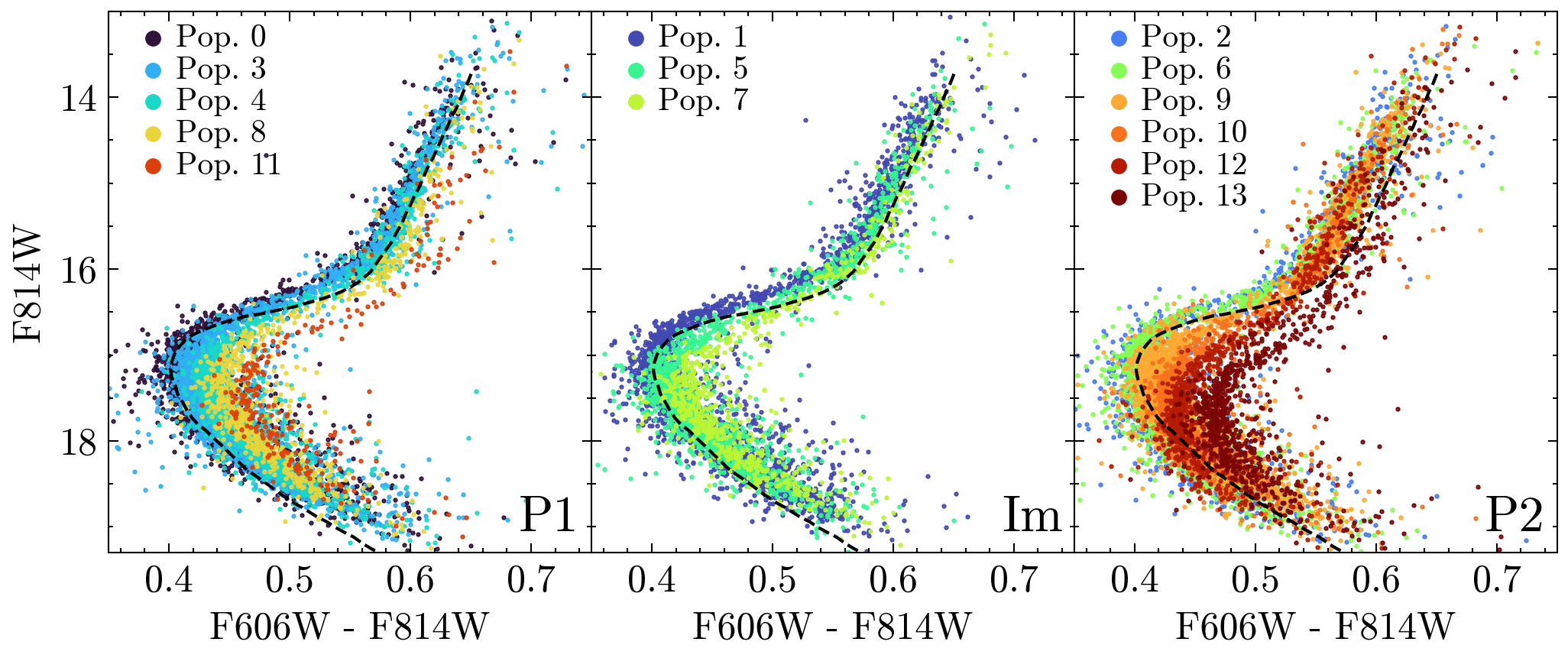}
    \caption{\textbf{Chromosome Diagram Grouped CMDs}: The F814W versus F606W - F814W CMD is shown for each set of populations grouped on the ChD (see Fig.\ref{fig:rgb_clustering}). The grouping (P1, Im, or P2) is annotation in the bottom right of each panel. Each data point is colored by its subpopulation assignment and indicated in the relevant legends. A fiducial line is given by a black dashed line, which traces the median metallicity population. Through comparison with the fiducial line we can confirm continuous sequences with CMD morphologies reflective of the expected variations in helium and alpha element abundances.}
    \label{fig:cmds_three_streams}
\end{figure*}

We show our subpopulation results across the full F814W range in Fig.\ref{fig:cmds_three_streams} where we separate the sequences into the P1, Im, and P2 groupings from the ChD for ease of visualization. Each grouping demonstrates not only the contiguousness of our subpopulations across the CMD, but the necessity for various helium abundance models to explain the spreads seen on the RGB and MS. For example, helium enhanced (Y = 0.40) isochrone models tend to have a less steep slope on the MS, a fainter MSTO, and a bluer RGB than their primordial helium (Y = 0.25) counterparts. This CMD morphological difference is especially highlighted when comparing the sharp MSTO of the metal-poor P1 populations to the metal-rich P2 populations. Similarly, varying alpha abundance models can account for spreads seen on the RGB that cannot be fully explained by metallicity differences. 

\begin{table*}[]
\centering


\begin{tabular}{cccccccccccccl}
\toprule

\multicolumn{6}{c}{\textbf{Subpop. Characteristics}} & \multicolumn{4}{c}{\textbf{Derived Parameters}} & \multicolumn{4}{c}{\textbf{Literature Comparison}}\\

\cmidrule(rl){1-6} \cmidrule(rl){7-10} \cmidrule(rl){11-14} 

Pop. & ChD. & AMR & $\rm{N_{*}}$ & \% & $\rm{N_{ages}}$ & [Fe/H] & $\sigma_{\rm{[Fe/H]}}$ & Age & $\sigma_{age}$ & B17 & B17 & L21 & D25 \\

\midrule
     0 & P1 &    Tight   & 6213 & 11.6 & 280 & -1.95 & 0.02 & 12.69 & 0.28 & rMS1 & 9.46\% & -1.68 & P1 \\
     1 & Im &    Tight   & 7877 & 14.7 & 405 & -2.01 & 0.04 & 12.21 & 0.33 & rMS2 & 9.71\% & -1.66 & $\rm{P2_{mid}}$ \\
     2 & P2 &    Tight   & 5661 & 10.6 & 266 & -1.90 & 0.04 & 11.87 & 0.54 & bMS1 & 13.0\% &-1.68 & $\rm{P2_{upper}}$ \\
     3 & P1 &  Diffuse   & 6007 & 11.3 & 295 & -1.81 & 0.03 & 12.50 & 0.32 & rMS3 & 14.0\% & -1.69 & AI \\
     4 & P1 &  Diffuse   & 3180 & 6.00  & 176 & -1.69 & 0.03 & 12.62 & 0.25 & rMS1 & 9.46\% & -1.68 & AI \\
     5 & Im &  Diffuse   & 3971 & 7.40  & 176 & -1.80 & 0.04 & 12.07 & 0.42 & MSe1 & 6.46\% & -1.52 & $\rm{AII_{mid}}$ \\
     6 & P2 &    Tight   & 5595 & 10.5 & 284 & -1.76 & 0.04 & 11.37 & 0.39 & bMS2 & 9.32\% & -1.54 & $\rm{AII_{mid}}$ \\
     7 & Im &  Diffuse   & 1525 & 2.90  &  93 & -1.63 & 0.05 & 12.08 & 0.59 & MSe2 & 6.56\% & -1.53 & $\rm{AII_{mid}}$ \\
     8 & P1 &  Diffuse   & 1751 & 3.30  &  89 & -1.58 & 0.05 & 12.54 & 0.09 & MSe3 & 1.17\% & -1.48 & AI \\
     9 & P2 &  Diffuse   & 5222 & 9.80  & 308 & -1.65 & 0.04 & 11.59 & 0.65 & bMS3 & 10.0\% & -1.43 & $\rm{AII_{upper}}$ \\
    10 & P2 &  Diffuse   & 2066 & 3.90  & 134 & -1.51 & 0.05 & 11.81 & 0.63 & MSd1,2 & 3.99\% & -1.26 & $\rm{AII_{upper}}$ \\
    11 & P1 & Metal-rich & 582  & 1.10  &  29 & -1.40 & 0.10 & 12.39 & 0.27 & MSe4 & 1.32\% &-1.43 & AI \\
    12 & P2 & Metal-rich & 1902 & 3.50  & 116 & -1.28 & 0.05 & 11.28 & 0.75 & MSd2,3 & 3.18\% & -1.18 & $\rm{AII_{upper}}$ \\
    13 & P2 & Metal-rich & 1804 & 3.40  & 101 & -1.04 & 0.05 & 10.70 & 0.69 & Msa1,2 & 3.53\% & -0.93 & $\rm{AII_{upper}}$ \\
\bottomrule
\end{tabular}
\caption{\textbf{Subpopulation Characteristics and Literature Comparison}: The Pop. column provides the number assigned to each subpopulation, ordered by metallicity. The ChD column provides the P1, Im, or P2 grouping of each subpopulation based on the ChD. The AMR column provides the Tight, Diffuse, or Metal-Rich grouping of subpopulations based on where they fall on the age-metallicity relation. The $N_{*}$ column provides the number of stars with a given subpopulation label. The $\%$ column indicates the percentages of the total labeled sample contained within a given subpopulation. The $\rm{N_{ages}}$ column reports for how many stars in a given subpopulation we have age constraints. The [Fe/H] column reports the median RGB metallicity of each subpopulation and the $\sigma_{[Fe/H]}$ column gives the intrinsic [Fe/H] spread. Similarly, the Age and $\sigma_{age}$ columns gives the median SGB ages and intrinsic age spread of each subpopulation. The final four columns show a correspondence of our subpopulations with the literature, more specifically \citealp{Bellini_2017c} (B17), where further details for the same populations are then given by \citealp{Latour_2021}, and \citealp{Dondoglio_2025}(D25) respectively. 
}
\label{table:subpops_details_and_lit_comp}
\end{table*}

\subsection{Verifying Subpopulation Consistency}
A critical aspect of this analysis is the connection of the RGB ChD to the SGB AMR. To ensure our connection between these two CMD regions is reliable, we compare their elemental abundance patterns using sodium abundances constrained by Wang et al., (in prep.) where they apply the DD-Payne machine learning algorithm \citep{Ting_2017, Xiang_2019, Wang_2022} to the MUSE spectra from \cite{Nitschai_2023} to obtain abundances for all well-measured stars. 

\begin{figure*}[htb]
    \centering
    \includegraphics[width = \textwidth]{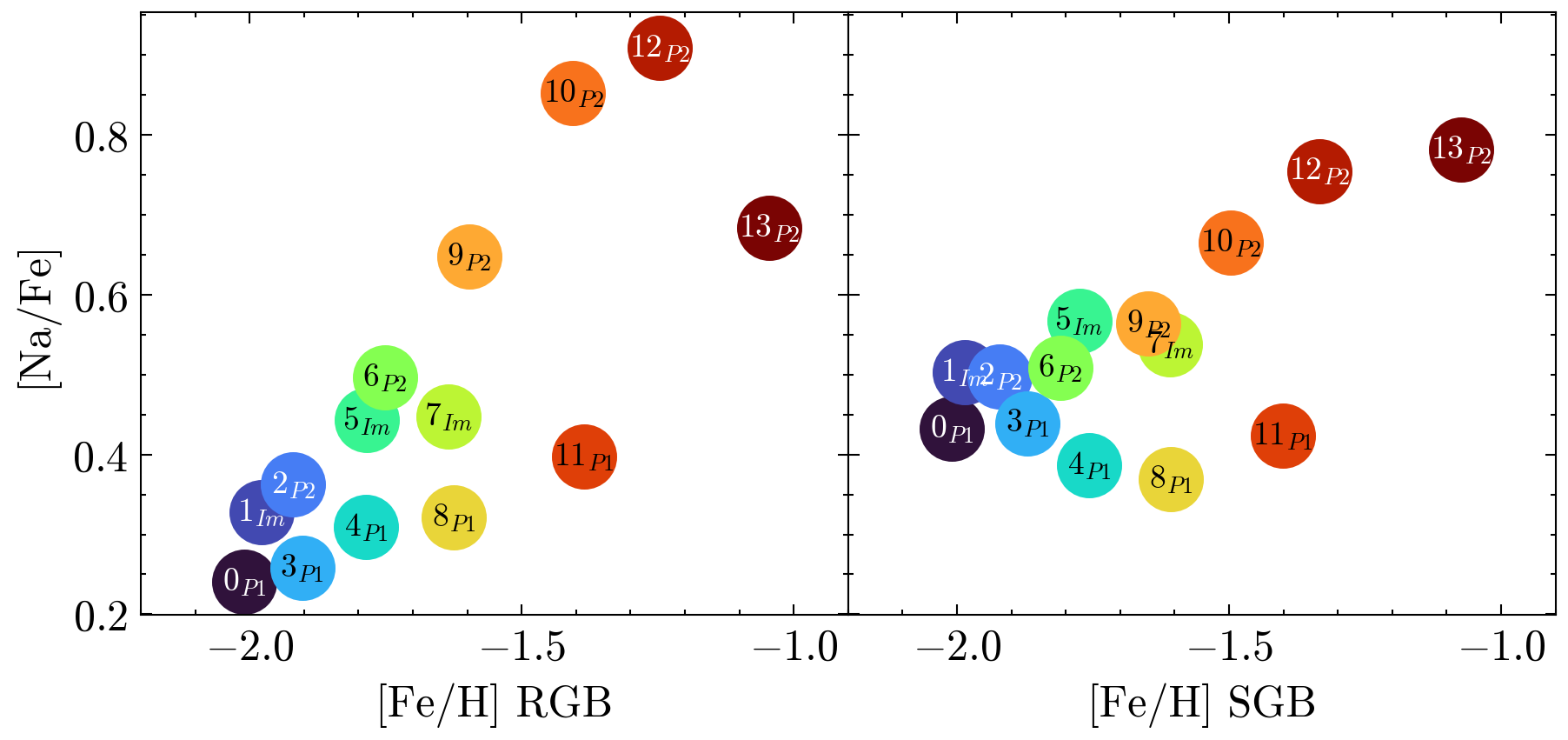}
    \caption{\textbf{[Na/Fe] Verification of Subpopulation Consistency}: \textit{(both panels)} Shown here is the median DD-Payne derived sodium abundance for the each subpopulations versus the respective median iron abundance. The left panel contains values for stars on the RGB (12 $<$ F814W $<$ 16 mag) and the right panel contains those for SGB stars (16 $<$ F814W $<$ 17 mag). We see consistency in the separation of P1, Im, and P2 across the panels with the overall range of sodium values decreasing on the SGB due to increased uncertainties.}
    \label{fig:na_fe}
\end{figure*}

Here, we use sodium abundances because they are found to show a large variations between the P1 and P2 populations and have relatively low median uncertainties (RGB: 0.07, SGB: 0.13). We first remove all stars observed with adaptive optics that show no flux in the NaID region of their spectra (as described in \cite{Wang_2025}) due to the sodium feature in these spectra being masked. We have a total of 6434 RGB stars and 5654 SGB stars with both Na abundances and cluster labels. We then calculate the median sodium abundance and median [Fe/H] for the RGB and SGB constituents of each population. The results are shown in Fig. \ref{fig:na_fe}. As expected, all populations show an increasing [Na/Fe] abundance with increasing metallicity. The P2 populations shown an enhancement in [Na/Fe] compared to P1 and the Im populations fall in between the two. This overall trend persists on the SGB, though the range of [Na/Fe] values is lower due to higher uncertainties in this region. We do see a shifting of populations 5 and 6, driven primarily by a slight change in their median metallicities, but in follow-up we confirm their distinction on the ChD at all magnitudes. This confirms the consistency in our subpopulation identifications across these groups and allows us to connect the RGB and SGB assembly probes with confidence. Lastly, we examined the spread in sodium abundances seen within each subpopulation. On the RGB the median 1-$\sigma$ range is 0.20 while on the SGB it is 0.26. In both cases this is 0.13 dex higher than the respective median uncertainties.

\subsection{Cross-match with SGB Ages}
\label{subsec:sgb_ages}

\begin{figure*}[htb]
    \centering
    \includegraphics[scale=.8]{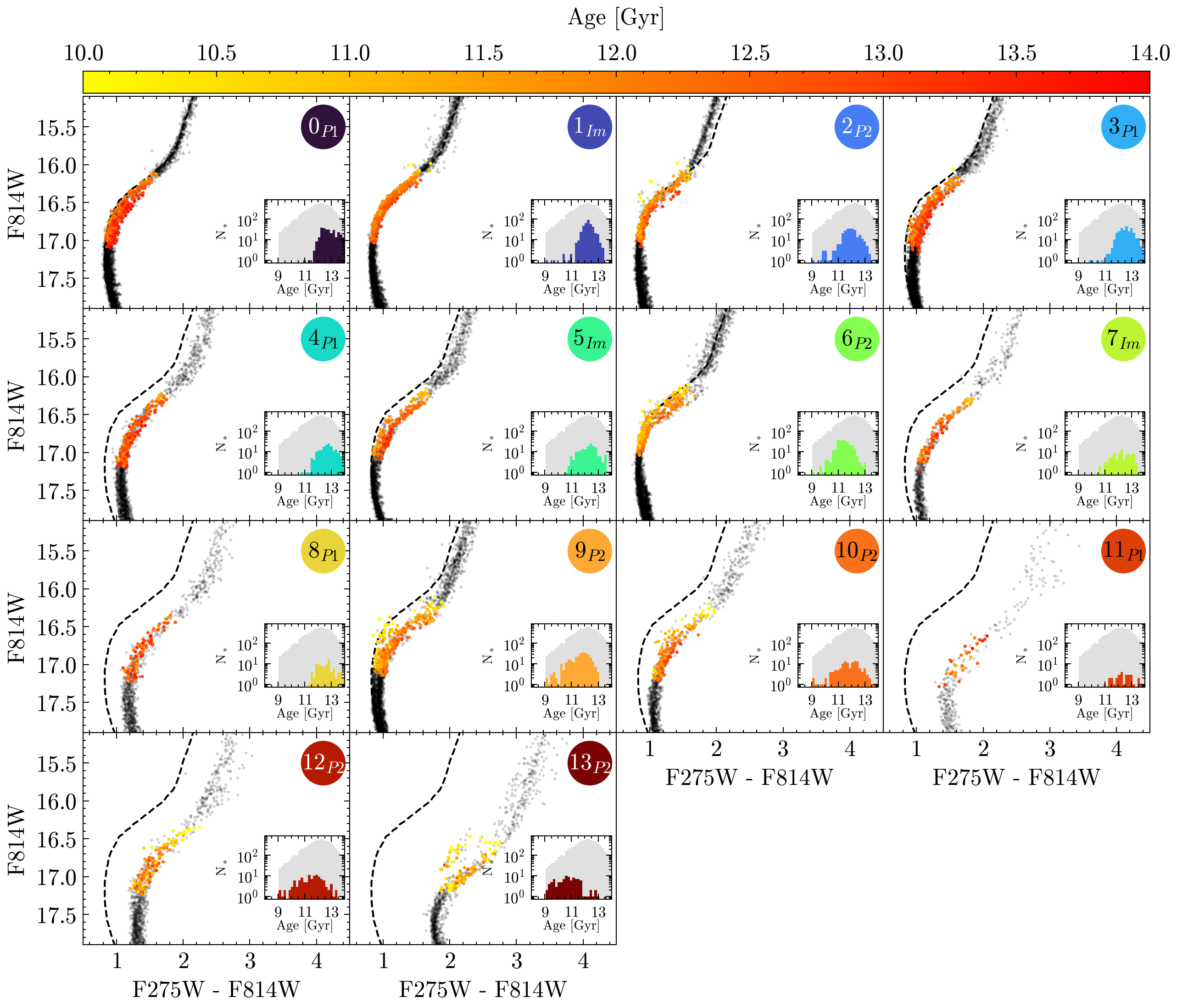}
    \caption{\textbf{Ages of Subpopulation Constituents}: One subpopulation is plotted within each subplot, top to bottom, left to right, from low to high metallicity. Within each panel, individual stars are plotted with black markers and SGB stars with age constraints are colored by their age. In the bottom right of each subplot there is an inset histogram showing in grey the full age distribution for all of our subpopulation labeled data and in the relevant color we shown the age distribution for that subpopulation. We also plot in each panel the same black dotted line which acts as a fiducial line for comparison across subpopulations.}
    \label{fig:subpops_w_sgb_ages_match}
\end{figure*}

Finally, matching our stars with those from the SGB ages work of \cite{Clontz_2025} gives us 2472 stars with both subpopulation cluster labels and reliable ages. The results are shown in Figure \ref{fig:subpops_w_sgb_ages_match} where we color each matched star on the SGB with its estimated age, given by the color bar along the top. We also plot the age histogram as an inset plot.

Looking at the age histograms, we see that many subpopulations have an overall age spread lower than the median age uncertainty reported in \cite{Clontz_2024} ($\sigma\sim$0.7 Gyr), suggesting our groupings are consistent with single-age/low age-spread stellar populations. Looking at the CMD sequences, we see the variety of MSTO and SGB morphologies across the populations reflect the expected variations in [$\alpha$/Fe] and helium abundances, highlighted by the deviations from the fiducial black dashed line drawn in each panel.

We also note in the last panel of Fig.\ref{fig:subpops_w_sgb_ages_match}, a cluster of stars in Pop. 13 which are brighter than the primary SGB sequence. One can also see them sitting slightly blue-ward on the SGB ChD (lower left panel) of Fig.\ref{fig:rgb_clustering}. It is clear they do not belong to either Pop. 11 nor Pop. 12. For now, we choose to leave them in Pop. 13, though we feel they are a separate population which is not fully distinguishable on the RGB. 

\subsection{Connecting Assembly Probes}
\label{subsec:connecting_assembly_probes}

\begin{figure}[htb]
    \includegraphics[scale=1.1]{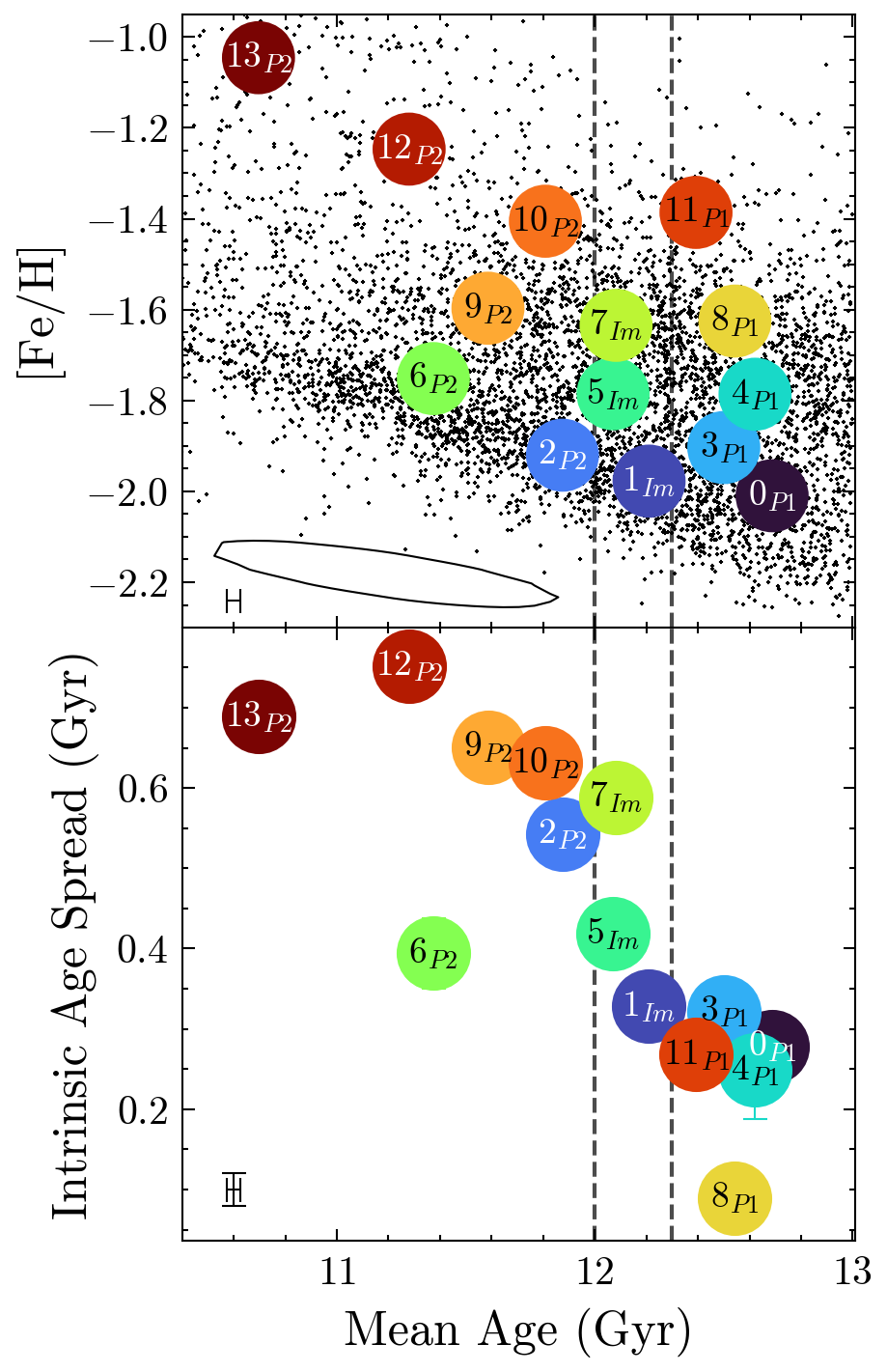}
    \caption{\textbf{Subpopulation Age Relations}: ($\textit{all panels}$) Each population is represented by a single medallion marker with the color corresponding to the cluster. Inside each the cluster number is annotated along with the relevant ChD stream indication subscript. The vertical black dashed lines show an approximate separation of the P1, Im and P2 populations. ($\textit{upper panel}$) Black points mark all SGB stars with ages from \cite{Clontz_2024}. Overplotted in the large colored circles are the mean (deconvolved) SGB age and median RGB metallicity for each subpopulation. The median age and [Fe/H] uncertainty for the SGB stars is given by the black contour while the median uncertainty on the mean age of each subpopulation is given by the black errorbars in the lower left. ($\textit{lower panel}$) Each subpopulation mean age is plotted versus its intrinsic age spread. The median uncertainty on the mean age and intrinsic age spread is represented by a set of errorbars in the lower left corner.}
    \label{fig:amr_and_age_spread_two_panel}
\end{figure}

Having a set of SGB stars with age constraints within each of our subpopulations means we can see in which region of the age-metallicity relation each subpopulation lies. To do this, we deconvolve the age histogram with the age uncertainty to calculate the mean age and intrinsic age spread of each subpopulation (following methods outlined in \cite{Clontz_2024}). The results are shown in the top panel of Fig. \ref{fig:amr_and_age_spread_two_panel}, where all stars with ages from \cite{Clontz_2024} are plotted in black points (with their uncertainties given by the black contour in the lower right) and each population is over-plotted with a large medallion marker at it's median RGB metallicity and mean SGB age. Each medallion population marker is labeled with the subpop. number and annotated with the relevant P1, Im, or P2 subscript. The median uncertainties in the mean ages are given by the deconvolution ($\sim$0.03 Gyr) and fall well within the extent of the medallions. For this reason we show the uncertainty on the mean age of the populations as a set of errorbars in the lower right corner. Looking at how the subpopulations occupy the AMR, we see the P1 populations stack vertically in this space, covering a large range in [Fe/H] but sharing similar ages of $\sim$12.5 Gyr. The same is true for the Im populations, sharing similar ages centered around $\sim$12.1 Gyr. The P2 populations are slightly more spread, with ages between 10.7 and 11.9 Gyr. This clearly shows that the P2 populations are all $>$ 0.80 Gyr younger than their similar-metallicity P1 counterparts and that the ages of the Im populations fall between those of P1 and P2 at all metallicities. This also shows that there is not a direct link between the 2-stream AMR sequences and the thre-stream feature of the ChD.

The mean age vs. the intrinsic age spread of each subpopulation is plotted with the same medallion markers in the lower panel of Fig. \ref{fig:amr_and_age_spread_two_panel}. The uncertainty on the age spread is only visible for Pop. 4, so we chose to represent the median uncertainty on both quantities as a set of errorbars in the lower left of this panel. The P1 populations cluster together at age spreads around 0.25 to 0.30 Gyr while nearly all P2 populations all have age spreads larger than 0.50 Gyr, reaching 0.75 Gyr for Pop 12. Pops. 6 and 8 are clear outliers, and it is notable that Pop. 6 is younger than the next two more metal-rich P2 populations, 9 and 10. The Im populations cover a large range of age spreads, from $\sim$0.3 to 0.6 Gyr, increasing with metallicity. Interestingly, they once again fall between P1 and P2 populations, suggesting they may be linked to these star formation events, instead of being brought in from GC inspiral, as suggested by \cite{Mason_2025}. All of the populations exhibit quite low age spreads overall ($<$0.1), consistent with globular clusters. In contrast, ages spreads seen in dwarf galaxy populations tend to be 0.30 - 0.40 Gyr \cite{Leaman_2012}. This is likely due to the built-in discreteness of our method. As mentioned, some of our subpopulations are likely segments of a larger continuous star formation episode.

\section{Discussion}
\label{sec:discussion}

In this section we discuss the implications of the connection of these subpopulations with the AMR in an effort to support or rule out potential formation mechanisms. We again emphasize that the main contribution this study adds is the age information. This does not allow us to definitively assert a formation model for the cluster but it does provide us with constraints which can support or rule out models which have been suggested in previous studies. We also examine the metallicity distributions of each population and the overall [Fe/H] distributions for the P1, Im, and P2 populations and to further contextualize our subpopulations, we compare their characteristics with those from the relevant literature.

\subsection{Subpopulations on the Age-Metallicity Relation}
In the upper panel of Fig.\ref{fig:amr_and_age_spread_two_panel} we show the connection between our subpopulations and the AMR. In \cite{Clontz_2024}, we suggest the shape of the lower/tighter AMR sequence is consistent with in-situ star formation, characterized by self-enrichment. This would suggest that Pops. 0, 1, 2, and 6, which all fall along the lower/tight AMR sequence (see Table \ref{table:subpops_details_and_lit_comp}), formed from successive star formation episodes, or perhaps they are segments of a continuous and extended star formation period proceeding in the same environment. The low age spread seen in Pop. 6 seems to suggest that population 6 experienced more rapid star formation. Looking at the time between these supposed successive star formation episodes we see that the age difference between Pops. 0 and 1 is 0.48 Gyr and between Pops. 1 and 2 is 0.36 Gyr, while the age difference between Pops. 2 and 6 is more than 1.31 Gyr. If we assume that populations that formed within a single environment follow similar slope tracks on the AMR, then we could theoretically also connect populations 3/5/9 and 4/7/10. Here we see age differences of $\sim$0.43 and 0.49 between 3, 5, and 9, respectively. Grouping 4, 7, and 10, we find age differences of 0.54 and 0.27 Gyr. After these three sequences, there exist no further Im populations, making it difficult to assert that there is a connection between P1, Im, and P2 across all metallicities. The age difference between Pops. 8 and 12 is 1.26 Gyr, and between Pops. 11 and 13 is 1.70 Gyr.

It is also worth noting that the overall intrinsic age spread vs. [Fe/H] relation seen in the lower panel of Fig. \ref{fig:amr_and_age_spread_two_panel} is similar to that shown in Fig. 5 of \cite{Clontz_2024}, where they constrain the age spread in bins of metallicity and find values of $\sim$0.4 $< \sigma < \sim$0.8 Gyr, with a similar flattening of the relation at the highest metallicities. In \cite{Clontz_2024} they also see a dip in overall age spread around [Fe/H] $\sim$-1.8 and -1.6, where here we see the low age spread contribution of Pops. 6 and 8, respectively, suggesting these low age spread populations are contributing to the dips seen in their metallicity-binned age spreads. 

\subsection{Metallicity Distributions of Subpopulations}
\begin{figure}[htb]
    \includegraphics[scale = 0.73]{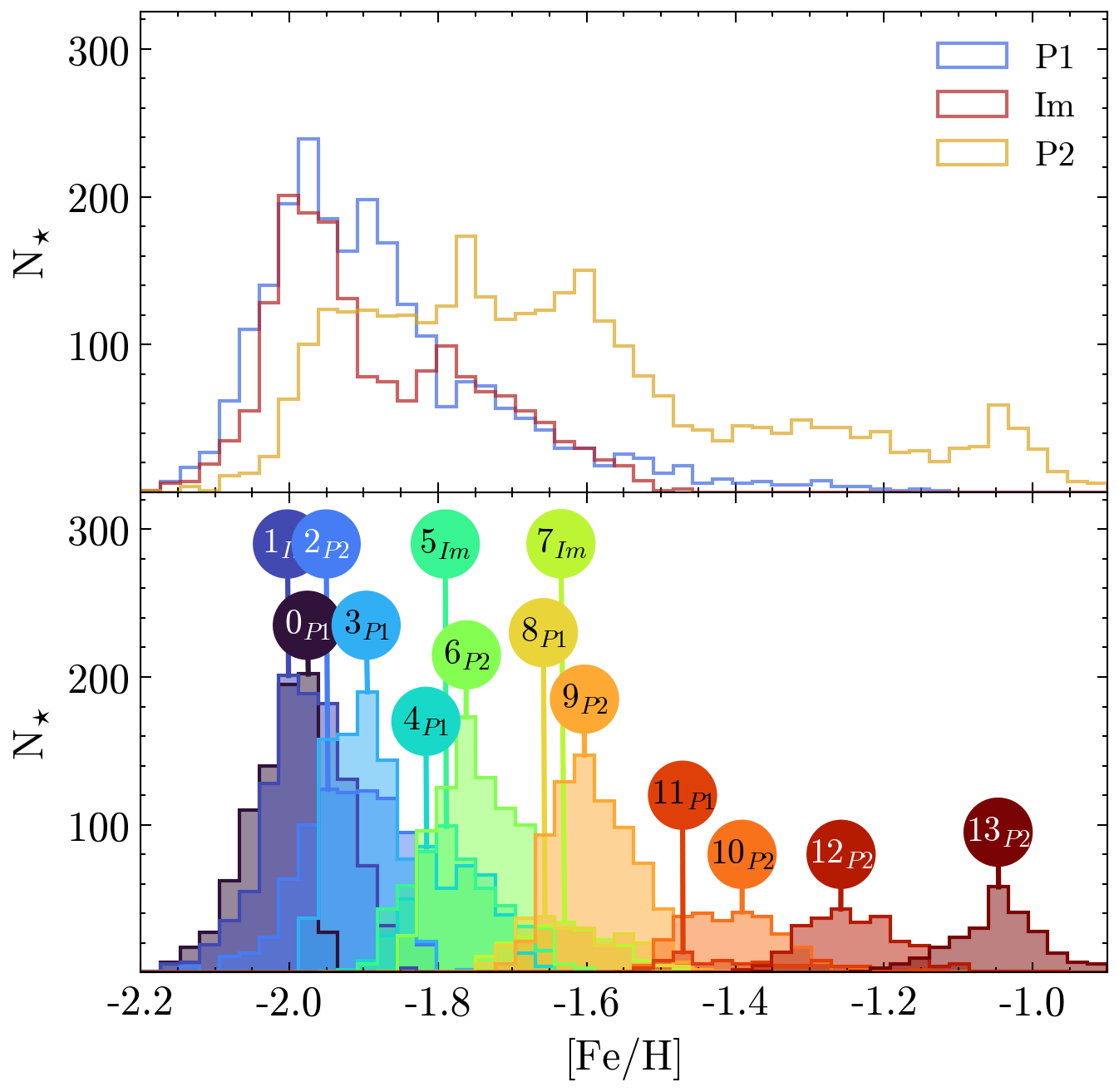}
    \caption{\textbf{Metallicity Distributions}: \textit{(upper panel)} The metallicity distribution functions of the full P1 (blue), Im (crimson), and P2 (gold) groups are shown. While the P1 and Im groups have similar distributions, that of P2 differs significantly. \textit{(lower panel)} The metallicity distribution functions of each of the individual subpopulations are shown, colored by their cluster number and labeled with the relevant medallion population markers at their peak value.}
    \label{fig:feh_histograms}
\end{figure}

The top panel of Fig. \ref{fig:feh_histograms}, shows the metallicity distribution functions of the grouped P1, Im, and P2 populations while the lower panel shows the metallicity distribution of each individual subpopulation. The P1 and Im populations show a remarkable similarity in their overall distributions suggestion a common origin or environment for these lower-metallicity subpopulations. P1 does see an additional peak around [Fe/H] -1.9, which clearly corresponds to population 3 in the lower panel. The Im population lacks stars above [Fe/H] of -1.5, while the P1 population has a small tail of stars extending to [Fe/H] = -1.1. Peaks are visible in the Im at [Fe/H] of -1.75 corresponding to Pop. 5. The P2 group has considerably fewer stars at low metallicities, compared to the P1 and Im groups. However, it has many more stars above [Fe/H] = -1.6, extending all the way to -0.9. There is considerable structure reflected in peaks corresponding to Pops. 6, 9, and 13. This extended metallicity distribution combined with the discreteness suggests perhaps a series of bursty star formation events within a single helium-enhanced environment; given the offset in metallicity and similarity in the light-element abundances, it seems perhaps this was separate from the environment where the P1 and Im stars formed.  

Alternatively, with the addition of the SGB age information from \citet{Clontz_2024}, the broader categories of P1, Im, and P2 could also be interpreted as the result of similar chemical enrichment in separate environments over a similar span of time as discussed in the previous subsection. If populations 0+1+2, 3+5+9, and 4+7+10 represent these similar enrichment patterns, it would suggest a relatively slow enrichment timescale ($\sim$1~Gyr) for light elements (helium/nitrogen) taking place in material with different starting metallicities over the same time period.  

\subsection{Comparison to Literature}
\label{subsec:comparison_to_literature}

To compare with previous studies of \omc's subpopulations, we discuss our 14 populations in terms of their average properties including their metallicity, their position on the ChD (P1, Im, P2), and the two-stream feature of the age-metallicity relation from \citet{Clontz_2024}. Due to the varying nature of how subpopulations are separated in the literature, we are limited to comparisons based on the relative qualities of the subpopulations.  

The only previous subpopulation identification of a comparable number of stars to this work are the 15 main sequence subpopulations sorted by \cite{Bellini_2017c} using an iterative selection on a variety of color-magnitude and color-color diagrams. Their work focused on the lower main-sequence and, therefore, the magnitude range of our classifications does not overlap enough for a direct comparison. However, we can connect our subpopulations through the chromosome diagram morphologies.


We can compare our RGB ChD to the MS ChD under the assumption that abundance variations (e.g. Nitrogen enhancement) have similar impacts on these the color of stars in this space. An inversion (flipping of the y-axis) of panel f of Fig. 10 in \cite{Bellini_2017c} indeed shows a similar morphology to that of our RGB ChD. In this comparison, the MSa connection with our subpopulation 13 is clear. The three bMS subpopulations: bMS1, bMS2, and bMS3 seem to correspond to our populations 2, 6, and 9, while the three MSd populations correspond to our populations 10 and 12. The three rMS populations cover both P1 and Im populations 0,1,3, and 4. The four MSe populations also cover P1 and Im populations 5, 7, 8, and 11. We also tabulate this comparison in Table \ref{table:subpops_details_and_lit_comp}. Another way to compare the populations is their fractional composition, which we also outline in Table \ref{table:subpops_details_and_lit_comp}. Lastly, \cite{Latour_2021} analysed MUSE observations for 4200 stars on the main-sequence and reported mean metallicities for each of the \cite{Bellini_2017c} subpopulations. We tabulate these as well. Where one of our populations is represented by a mixture of two populations from \cite{Bellini_2017c} we list the average.

Recently, \cite{Mason_2025} used spectroscopic abundances to separate populations by applying Gaussian Mixture modeling to a multi-dimensional abundance space using $\sim$1500 APOGEE VAC stars belonging to \omc\ and were able to identify three groupings. These roughly translate to the P1, Im, and P2 populations on the ChD (see their Fig. 8). Their claim that the Im populations have a separate chemical enrichment consistent with globular clusters is well supported by the abundance patterns seen, especially in [Al/Fe] vs. [Mg/Fe] space (see their Fig. 5). Based on the interpretation of the population histories in the \citet{Mason_2025} paper, this would suggest that our Pops. 1, 5, and 7 are NSC assembly contributions from GC in-spiral. While certainly possible, we note that the Im populations fall between P1 and P2 in every probe we examine (age, age spread, and sodium abundance), which would require a perhaps unlikely coincidence between these quantities in different formation environments. They also run chemical evolution models from which they conclude P1 formed first, in the center of \omc's host galaxy and that P2 formed later, perhaps with GC origin. Our age constraints are not inconsistent with these models. They again note the mass budget problem as an added complexity to constraining formaiton models for these groups.

The recent publication by \cite{Dondoglio_2025} discusses subpopulation in \omc\ in the context of the ChD and APOGEE abundances covering out to 5 half-light radii. They separate their RGB stars into three streams in the ChD and then further distinguish subgroupings in each stream. In the lower stream, they identify a first-generation (1P) population and an anomalous population (AI). For several of their analyses the middle and upper stream are combined, but are also separated into a metal-poor second-generation (2P) and additional more metal-rich anomalous population (AII). The correspondence with the subpopulations form this work is outlined in Table \ref{table:subpops_details_and_lit_comp}. In their Table C.1 they outline their formation scenario for each grouping and how they fit together. Their primary finding is that models can, in the case of aluminum abundance trends with metallicity, explain the formation of the middle stream stars as having formed from 2P ejecta diluted with 1P-like-gas. When comparing with this work, the main difference is our age constraints, which show the middle stream stars (Im populations, here) formed between the lower and upper stream stars in time. This argues against their dilution formation mechanism for the Im population. Another interesting finding is that the 2P stars are more centrally concentrated than the 1P stars and this is mirrored in the AI and AII populations respectively. This supports the idea that the P2/2P stars formed from gas that was enriched by P1/1P ejecta that then sank to the center of the cluster before forming P2. This formation model implies all of \omc's populations formed through self-enrichment processes, a model which struggles to reproduce the differences in metallicity distributions between P1 and P2 groupings. 

All of these studies represented several significant steps forward in understanding the formation of \omc. Additional theoretical work on polluter mechanisms and the mass budget problem will surely soon uncover the complexity of its assembly.

\subsection{Numbers of Subpopulations, their Discreteness, and their Origin}

In this study we find 14 subpopulations that exist across the RGB, SGB, and MSTO. The fact that we detect a similar number to what was previously found by \cite{Bellini_2017c} on the lower MS, using different photometric analyses, is strong evidence for the number of identifiable subpopulations to be around 15. While undoubtedly some of these individual subpopulations or combinations of these subpopulations correspond to discrete evolutionary events or locations, some may also be part of a single formation channel spread out over time, as we suspect for the lower stream of the age-metallicity relationship. For example, the low age spread in population 6 and 8, and the very distinct age and metallicity of population 13 may correspond to a discrete formation event or environment separate from the rest of the cluster. It is clear that as more and better data become available, our ability to resolve the origin of these subpopulations will also improve. In particular, high-precision abundances across these subpopulations can enable determinations of the discreteness or continuity of star formation events among subpopulations. 

\section{Conclusions} 
\label{sec:conclusions}

This study combines multi-band photometry with precise spectroscopic metallicities that together provide, for the first time, the constraining power necessary to parse the multiple subpopulations of \omc\ into individual sequences from the upper RGB to the upper MS. 

This analysis immediately provides information on the number of identifiable populations and their relative number densities. In addition, it allows, for the first time, the connection between the generations of stars (identifiable on the ChD) and the multiple streams seen in the AMR. This shows definitively that the two-stream feature of the AMR does not map directly to the three-stream feature of the ChD. Exactly how to group the populations on the AMR is still unclear. The combination of these results with the DD-Payne abundances from Wang et al. (in prep.) will provide additional insight into the star formation history and assembly of \omc. 

Thus, from the analyses outlined above we have concluded upon the following key points.

\begin{itemize}
    \item In \omc\ we identify $\sim$14 subpopulation groupings on the RGB down to below the MSTO. This likely does not represent the full set of distinct stellar populations that constitute this cluster but serves as a benchmark for further studies.

    \item The populations that fall into each of the P1, Im, and P2 groupings have similar mean ages across all metallicities, with the P1 populations being oldest at $\sim$12.6 Gyr, the Im populations having mean ages about 0.4 Gyr younger than the P1 populations, and the P2 populations all having mean ages around 11.6 Gyr. The median age of the P2 population is 1.1 Gyr younger than the median age of P1. 

    \item The P1 populations show lower intrinsic age spreads ($\langle \sigma_{intrins.} \rangle= 0.24$ Gyr) compared to the P2 populations ($\langle \sigma_{intrins.} \rangle = 0.61$ Gyr). The Im population shows a large range of age spreads, falling in between the P1 and P2 groupings.

    \item One interpretation of population occupation of the AMR is that sets of P1, Im, and P2 populations trace diagonal lines, most notably within the distinct lower stream found by \citet{Clontz_2024}. This could suggest parallel evolution and light element enrichment in multiple environments at the same time. Another interpretation is that there are three relatively vertical sequences in the AMR, traced by P1, Im, and P2 separately. In this model the connection with the structure of the AMR complicated.

\end{itemize}

Despite this wealth of information, a fundamental puzzle of the origin of \omc\ and its unique stellar populations remains. Formation must have taken place in discrete environments, but how the evolution of the different subpopulations relate to one other remains uncertain. Several ongoing and planned works will continue to add detail to the model of \omc's assembly. Follow up high-resolution spectra observed earlier this year with FLAMES/GIRAFFE@VLT will soon provide additional constraints on the different formation mechanisms for the two sequences of the AMR and ongoing work studying the abundance differences between the most-metal-rich P1 and P2 populations will help us understand the very anomalous Population 13. Additionally, updated isochrone models that can reproduce the F336W magnitudes of the nitrogen enhanced stars are generated and will be crucial for allowing isochrone fitting and proper synthetic stellar population modeling of the enhanced populations of this cluster. A deeper look into the mean abundances of light elements in each subpopualation is underway, using a machine learning methods on the MUSE spectra (Wang et al., in prep.).




\begin{acknowledgments}

CC acknowledges the contributions to this work via the high performance computing resources at the University of Utah as well as the cluster computing resources of the Max-Planck Institute for Astronomy Heidelberg. ACS, ZW and CC acknowledge support from a Hubble Space Telescope grant GO-16777. M.A.C. acknowledges the support from FONDECYT Postdoctorado project No. 3230727. AFK acknowledges funding from the Austrian Science Fund (FWF) [grant DOI 10.55776/ESP542]. The authors thank Davide Massari and Ricardo Schiavon for helpful discussions.

\end{acknowledgments}

%

\vspace{5mm}
\facilities{HST(STScI), MUSE@VLT()}


\software{Numpy \cite{numpy}, Matplotlib \citep{matplotlib}, Astropy  \citep{astropy:2013, astropy:2018, astropy:2022}, Scikit-Learn \citep{scikit-learn}}

\appendix
In Appendix \ref{appendix:algorithm_variation_testing} we describe in more detail the additional tests which were performed to finalize our algorithm methodologies outlined in Section \ref{sec:methods}.

\section{Algorithm Variation Testing}
\label{appendix:algorithm_variation_testing}

With regards to our sample selections in Section\ref{sec:step_1_sample_selection}, for several of our quality cuts we experimented with various thresholds, with the goal of balancing sample size with sample quality. Future clustering steps were found to be strongly affected by outliers in [Fe/H] and to rely heavily on this dimension at fainter magnitudes. Therefore, the above criterion of [Fe/H] uncertainty $<0.3$ was required. Additionally, we tested our clustering algorithm performance when using the DD-Payne iron abundances calculated in Wang et al. (in prep.), using [M/H] instead of [Fe/H], as well as excluding [Fe/H] all together and did not find that any of these provided more pristine and/or distinct subpopulations than using the [Fe/H] values derived from the MUSE metallicities.

In Section\ref{sec:step_2_prep_clustering_dimensions} where we prepare our data's clustering dimensions, we also tested several iterations of varied fiducial line choices, including metallicity-dependent fiducial lines, the non-normalization of the colors, and principal component analysis, with no significant improvements made to the tracking of the subpopulations to fainter magnitudes. We additionally tested the performance of our clustering algorithm when using all six available photometric bands, which tended to obscure the subpopulations more and often caused the propagation algorithm to behave in unpredictable ways. Thus, we chose to keep our analysis to the three dimensions listed in the main text, the two delta colors \deltaone and \deltatwo, plus [Fe/H].

When we did the $N_{subpops}$ determination in Section\ref{sec:step_4_n_subpops_determination}, 
we also note that this number varies (from $\sim$12 to $\sim$17) based on the scaling of dimensions and the total number of RGB stars considered. However, we find there are considerable disadvantages to fewer or more clusters in later steps of our modeling. Fewer clusters lead to groupings with two or more distinct (by-eye) overdensities on the chromosome map being grouped together, often causing issues with the averaging of their metallicities complicating propagation. With more clusters, we can further distinguish groups that are clearly separate on the RGB. However, these groups are substantially overlapping in color at fainter magnitudes, making it very challenging to separate them with confidence. We performed several checks of clustering performance through the testing of literature subpopulation numbers, such as the 15 found by \cite{Bellini_2017c}. The primary difference here is the separation of the two most metal-rich populations, which for us are grouped together (in Pop. 13). We find the ``stealing" by Cluster 13 from the next metal-poor population (12 in our algorithm) is minimized when we do not split Cluster 13 into two. 

Additionally, we tested our ability to isolate and track 3 primary populations (P1, Im and P2) given by the three streams on the RGB ChD, which proved to be messy below the RGB. Given these issues and others, we chose 14 subpopulations for this analysis, but do not assert that this is strictly the absolute number of subpopulations in \omc\ nor that each is a pristine single stellar population. 

During the assembly and testing of our current RGB clustering algorithm in Section\ref{sec:step_5_rgb_clustering}, we also tested outcomes using a Gaussian Mixture Model as well as a Bayesian Gaussian Mixture Model as the algorithm which provided initial cluster labels to the RGB. The primary drawback of these methods is that they require a Gaussian shape for each over-density. This would cause some clusters, such as Cluster 11 in the current method, to be extended in all clustering dimensions, causing a large range of [Fe/H] values to be assigned. This method also always produces a background term, a low-amplitude high $\sigma$ term meant to characterize the noise within a sample. This term would need to be removed before each step of propagation, and because of the already strict outlier removal processes, felt unnecessary. These Gaussian mixture models are also typically built on the assumption that each cluster has approximately the same variance, which is certainly not the case, given the extent of e.g. Cluster 11 vs. Cluster 4. 

Lastly, in Section\ref{sec:step_7_propagation}, where we begin to propagate our subpopulation labels, we would like to mention that this approach was preceded by many other methods, each of which had its advantages and disadvantages, and none of which performed better than our current algorithm when it came to keeping the color spread and metallicities consistent within each given subpopulation across the CMD. We tried the NearestCentroid algorithm \citep{scikit-learn}, assigning each star to its closest centroid, which led to the confusion of populations that cross around the MSTO. This also, in some cases, led to wandering cluster centers when clusters had a large variance. We additionally tested a KNearestNeighbor \citep{scikit-learn} approach where the e.g. 10 nearest-labeled stars to a given unlabeled star voted on which label the target star received. This led to some clusters growing disproportionately large in a cascade of overassignment as we propagated to fainter magnitudes. 

We also tested using the cluster assignments as training data for both a Gaussian Mixture Model and a K-Means clustering algorithm, which in theory would simply find the same $\rm{N_{clusters}}$ with similar centroids and variances in the next faintest bin. However, the initial issue was that the cluster numbers were random and would need to be resorted at every metallicity, which could cause populations with similar metallicities to become mixed with one another. Even after solving this, the re-seeding of the subpopulation centroids, even with strong priors, did not produce consistent results across the CMD. 

\section{Data Availability}
\label{appendix:data_table}

 The results of our subpopulation analysis are provided as a machine-readable table \textbf{to be added upon acceptance}. The columns provided in this table are outlined in Table \ref{table:catalog_columns}.
 
\begin{table}[]
    \centering
    \begin{tabular}{l|l}
    \scriptsize
    Column Name & Description \\
    \hline
    HST ID  &  \cite{Haeberle_2024} Catalog Identifier  \\
    MUSE ID &  \cite{Nitschai_2023} Catalog Identifier \\
    delta\_275\_814 & $\Delta$ color for F275W - F814W   \\
    delta\_275\_336\_435 & $\Delta$ color for (F275W - F336W) - (F336W - F435W) 
    \\
    Subpopulation Tag &  Integer indicating cluster assignment \\
    \hline
    \end{tabular}
    \caption{Provided Subpopulation Catalog: Each column of the provided catalog is described. All quantities are unitless.}
    \label{table:catalog_columns}
\end{table}

\bibliography{main_bib}
\bibliographystyle{aasjournal}



\end{CJK*}
\end{document}